\documentclass[%
 reprint,
superscriptaddress,
amsmath,amssymb,
aps,
pra,
]{revtex4-1}

\usepackage{graphicx}
\usepackage{dcolumn}
\usepackage{bm}
\usepackage{xcolor}
\usepackage{upgreek}
\usepackage{lineno}
\usepackage{mathtools}
\usepackage{siunitx}

\usepackage{mathtools}
\newcommand{\ko}{k_{\mathrm{o}}}
\newcommand{\e}{\mathrm{e}}

\newcommand{\M}{\mathrm{M}}
\newcommand{\B}{\mathrm{B}}
\newcommand{\A}{\mathrm{A}}

\newcommand{\kL}{k_\mathrm{L}}
\newcommand{\ks}{k_s}
\newcommand{\zs}{z_s}
\newcommand{\Jzero}{J_0}

\begin{document}

\preprint{APS/123-QED}
\title{Self-healing of the Montgomery pattern}

\author{Athena Xu}
\affiliation{Harvard John A. Paulson School of Engineering and Applied Sciences, Harvard University, Cambridge, MA, USA}
\affiliation{Department of Electrical and Computer Engineering, University of Waterloo, Waterloo, N2L3G1, ON, Canada}

\author{Oscar de Vries}
\affiliation{Harvard John A. Paulson School of Engineering and Applied Sciences, Harvard University, Cambridge, MA, USA}
\affiliation{Department of Applied Physics and Science Education, Eindhoven University of Technology, 5600 MB Eindhoven, The Netherlands}

\author{Alfonso Palmieri}
\affiliation{Harvard John A. Paulson School of Engineering and Applied Sciences, Harvard University, Cambridge, MA, USA}

\author{Murat Yessenov}
\thanks{Corresponding authors: yessenov@seas.harvard.edu}
\affiliation{Harvard John A. Paulson School of Engineering and Applied Sciences, Harvard University, Cambridge, MA, USA}
\affiliation{CREOL, The College of Optics \& Photonics, University of Central Florida, Orlando, Florida 32816, USA}

\author{Ayman F. Abouraddy}
\affiliation{CREOL, The College of Optics \& Photonics, University of Central Florida, Orlando, Florida 32816, USA}

\author{Federico Capasso}
\thanks{capasso@seas.harvard.edu}
\affiliation{Harvard John A. Paulson School of Engineering and Applied Sciences, Harvard University, Cambridge, MA, USA}

\begin{abstract}
Self-healing -- the ability of a structured beam to reconstruct its transverse profile after partial obstruction -- has been demonstrated for diffraction-free beams, where the recovery distance varies continuously with obstruction size. Here, we investigate self-healing in the Montgomery pattern, a self-imaging of tightly localized optical fields. Using Babinet's principle, we show theoretically that the recovery distance is quantized in integer multiples of the self-imaging period -- a qualitative distinction from all previously studied self-healing beams. We confirm these predictions experimentally using a programmable holographic setup with circular disk obstructions of size up to $20\times$ of the spot size of the Montgomery pattern at the self-imaging plane, establishing the robustness of the Montgomery pattern against scatterers and obstructions in the beam path.
\end{abstract}


\maketitle

Self-imaging -- lens-less revival of the wave field at discrete axial planes -- has been a topic of interest in optics since its first observation by Talbot~\cite{Talbot36PM}. Whereas the Talbot effect relies on the periodic sampling of the transverse profile and yields a fractal tapestry of intensity across the propagation volume, the Montgomery effect reveals that revivals persist under a far broader class of aperiodic sampling, provided the axial wavenumber is linearly discretized~\cite{Montgomery67JOSA,jahns_montgomery_2003, Indebetouw1992JOSAA}. For cylindrically symmetric fields, this condition translates into a square-root discretization of the radial spatial frequency, known as Montgomery rings \cite{lohmann_fractional_2005}, which produces periodic axial revival of tightly localized spots. The spatially structured Montgomery effect was recently observed in free space using a dynamic phase hologram, demonstrating independent control of the focal-spot size, depth of focus, and self-imaging distance, as well as the arbitrary shape of transverse profile~\cite{Yessenov26Optica}.  These features make Montgomery self-imaging attractive for plane-selective manipulations in atomic systems ~\cite{Kusano2025PhysRevRes}, three-dimensional optical trapping~\cite{Schlosser2023PRL} for a large array of neutral atom traps \cite{chiu2025Nature}, ring traps for quantum gases~\cite{De2021JoP,Ryu13PRL}, and simultaneous multi-plane or volumetric imaging~\cite{Chen2024NatMet}. 

In this Letter, we ask whether the Montgomery self-imaging fields possess self-healing properties analogous to those of diffraction-free beams. Self-healing -- the ability of a beam to restore itself after partial obstruction -- has been studied primarily in the context of diffraction-free beams, such as Airy beams~\cite{Siviloglou07OL, Broky08OE}, Bessel beams~\cite{Durnin87PRL, Bouchal1998OC, chu_analytical_2012,Aiello14OL}, and space-time wave
packets~\cite{Kondakci17NP,Yessenov19PRA,Kondakci18OL}. Although both self-imaging and diffraction-free properties of light stem from structuring light fields \cite{Yessenov22AOP,Shen2023Roadmap}, the underlying mechanisms resulting in these two effects are fundamentally distinct. Diffraction-free beams continuously maintain their spatial profile over an extended distance, whereas the self-imaging fields reappear at discrete axial planes. 

Here, we investigate the self-healing of the Montgomery pattern theoretically and demonstrate experimentally. Treating the Montgomery field as a coherent superposition of Bessel beams with square-root spacing of radial spatial frequencies, we find that the intensity profile of the Montgomery pattern consists of two parts with distinct properties: a propagation-invariant term and a self-imaging term. Using this analytical model of the Montgomery pattern, we investigate the self-healing distance for a partial obstruction placed at the self-imaging plane. We find a qualitative distinction from conventional self-healing beams: because the Montgomery field reconstructs only at discrete self-imaging planes, its recovery distance is quantized in integer multiples of the Montgomery self-imaging period. In the single-ring limit, the theory reduces to the familiar Bessel-beam result, whereas for multiple coherent rings the recovery depends jointly on the obstruction geometry and the spectral power distribution across the Montgomery rings. We verify these predictions using a holographic optical synthesis setup for the Montgomery pattern and direct measurements of its reconstruction behind an opaque obstruction. We use a circular disk obstruction of various radii from 50~\unit{\um} to 250~\unit{\um}, which is up to $20\times$ of the beam size at the self-imaging plane, and record the intensity profiles of obstructed Bessel beam, Montgomery patterns with self-imaging distances from 4~mm to 10~mm, and beam sizes from $10$~\unit{\um} to $18$~\unit{\um}. We demonstrate that the self-healing distance of a Bessel beam is continuous, whereas for the Montgomery pattern it is discretized. To the best of our knowledge, this is the first demonstration of the self-healing of Montgomery patterns, which exposes its robustness against scatterers and apertures, a feature that has applications in
simultaneous micro-manipulation of particles in multiple planes~\cite{garces-chavez_simultaneous_2002}, light-sheet microscopy in inhomogeneous or scattering media~\cite{Fahrbach10NP}, and, more recently, image encryption with Bessel--Gauss beams~\cite{baliyan_unveiling_2025}.

\begin{figure}[t!]
\centering
\includegraphics[width=8.6cm]{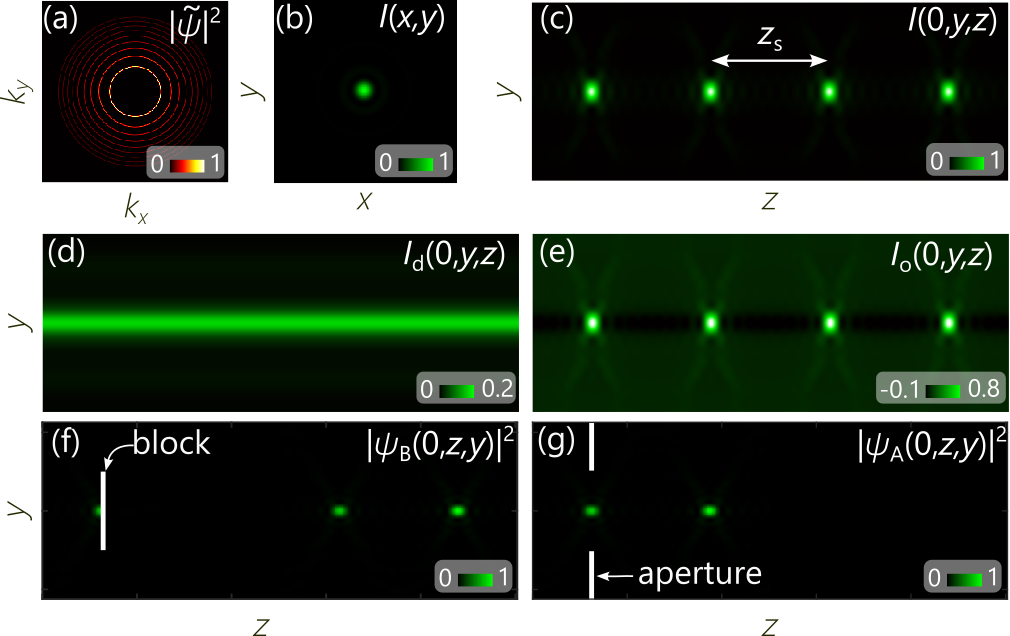}
\caption{The concept of the Montgomery effect and its self-healing mechanism. (a) The spatial spectrum $|\widetilde{\psi}(k_x,k_y)|^2$ of the field underlying the Montgomery pattern, (b) transverse intensity profile $I(x,y)$ at the self-imaging plane $z\!=\!z_s$, and (c) the axial intensity profile $I(y,z)$ at $x\!=\!0$. (d) The diagonal $I_d(y,z)$ intensity contribution possesses a diffraction-free behavior, while (e) the off-diagonal term $I_o(x,z)$ self-images at discrete planes $z\!=\!qz_s$, note that $I_o\!<\!0$ around $y\!=\!0$, $z\!\neq\!\zs$. (f) The intensity evolution of the Montgomery pattern $|\psi_{\B}(r,z)|^2$ with an obstruction placed at $z\!=\!z_s$ and (g) corresponding complementary apertured intensity profile $|\psi_{\A}(r,z)|^2$. }\vspace{-5mm}
\label{Fig:Concept}
\end{figure}


 We first outline the theoretical basis for the Montgomery pattern. Throughout, we consider scalar paraxial monochromatic fields of frequency $\omega_{o}$ (wavenumber $k_o=\omega_o/c$) propagating along the $z$-axis, where $c$ is the speed of light.  Writing the field in cylindrical coordinates $(r,\varphi,z)$ as $E(r,\varphi,z;t)=e^{i(k_{o}z-\omega_{o}t)}\,\psi(r,\varphi,z)$, the envelope is expressed by
 \begin{equation}
 \begin{split}
          &\psi(r,\varphi,z)\!=\\
          &\!\!\int_{0}^{\infty}\!\!\int_{0}^{2\pi}\!k_{r}\widetilde{\psi}(k_{r},\chi)\,e^{ik_{r}r\cos(\varphi\!-\!\chi)}e^{\!-i(k_{o}\!-\!k_{z})z}\,d\chi\,dk_{r},
 \end{split}
 \end{equation}
where $\widetilde{\psi}(k_r,\chi)$ is the spatial spectrum, $k_z$ and $k_r$ are the axial and radial spatial frequencies, respectively, and $\chi\!=\!\arctan{(k_y/k_x)}$, and  $k_x,k_y$ are the transverse spatial frequencies in Cartesian coordinates. Throughout this paper, we consider optical fields in the paraxial regime for which $k_{z}\!\approx\!k_{o}\!-\!k^{2}_{r}/2k_{o}$. 
Self-imaging at discrete planes $z=qz_s$ ($q\in\mathbb{Z}^{+}$) requires $(k_o-k_z)z_s=2\pi n$ ($n=1,2,\hdots N \subset\mathbb{Z}^{+}$), which leads to Montgomery self-imaging condition \cite{Yessenov26Optica, Montgomery67JOSA}: $k_z(n)=\ko-k_sn,\qquad k_r(n)=\kL\sqrt{n}.$
Here $n$ is the ring index, $N<\zs/\lambda_o$ is index of the largest ring, $\ks=2\pi/z_s$ and $\kL=\sqrt{2\ko\ks}$ are sampling coefficients in $k_z$ and $k_r$ respectively [Fig.\ref{Fig:Concept}(a)]. Within this work, we focus on cylindrically symmetric fields, for which the spatial spectrum of an ideal Montgomery field can be represented as 
$\widetilde{\psi}(k_r)=\sum_{n=1}^{N}\widetilde{c}_n\delta(k_r-\kL \sqrt{n})$,
which corresponds to concentric rings in the spectral space, known as Montgomery rings; here $\widetilde{c}_n$ is the complex coefficient of each Montgomery ring and $\delta(\cdot)$ is the Dirac delta function. The corresponding field profile in the physical space takes the form 
\begin{equation}
    \psi(r,z)\propto\sum_{n=1}^{N}c_{n}\Jzero^{(n)}(r)\e^{-i\ks nz},
\end{equation}
where $\Jzero$ is the zeroth-order Bessel function of the first kind [Fig.~\ref{Fig:Concept} (b-c)]. Each Montgomery ring of index $n$ in the Fourier space corresponds to a Bessel beam $\Jzero^{(n)}(r)=\Jzero(\kL\sqrt{n}r)$ with a complex weight coefficient $c_n\!\propto\widetilde{c}_n \sqrt{n}$ and propagation angle $\theta_{n}=\tan^{-1}\left[ \frac{\kL\sqrt{n}}{\ko-k_s n}\right]$. After some algebra, the intensity profile of the Montgomery pattern can be represented as a sum of diagonal and off-diagonal terms (see \cite{Supplementary} for more details)

\begin{figure}[t!]
\centering
\includegraphics[width=8.6cm]{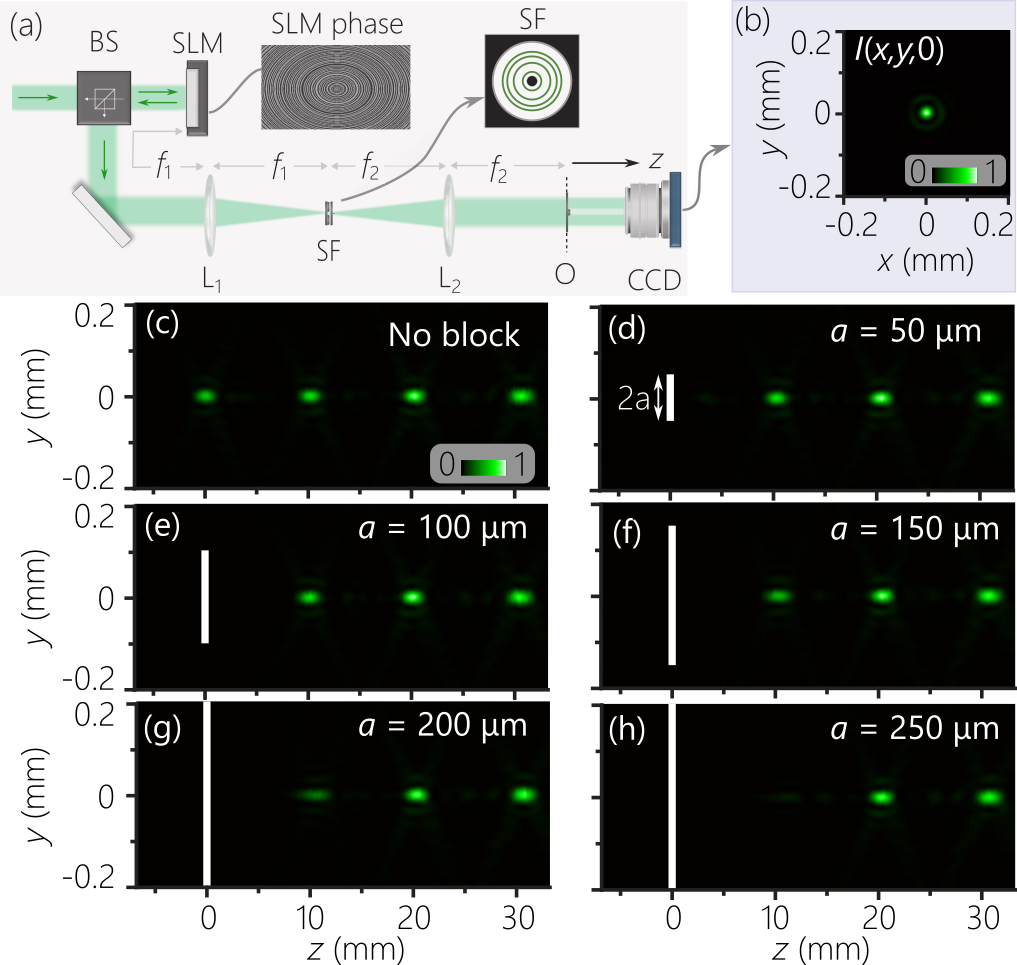}
\caption{(a) Experimental setup for synthesis and characterization of the Montgomery pattern. SLM: spatial light modulator; BS: beam splitter, SF: spatial filter; L: spherical lens; O: obstruction (b) Measured transverse intensity profile $I(x,y)$ recorded at $z=0$ on the CCD for the Montgomery pattern of self-imaging distance $z_s\!=\!10$~mm and $N\!=\!7$. (c) Axial intensity evolution $I(0,y,z)$ of the Montgomery pattern with no block. (d) Measured Montgomery pattern behind a circular disk obstruction of radius $a\!=\!50$~\unit{\um} placed at $z=0$. (e-h) same as (d) for obstruction radii of $a\!=\!100$~\unit{\um}, $150$~\unit{\um}, $200$~\unit{\um} and $250$~\unit{\um}, respectively.}\vspace{-5mm}
\label{Fig:Fig2}
\end{figure}

\begin{figure*}[t!]
\centering
\includegraphics[width=17.6cm]{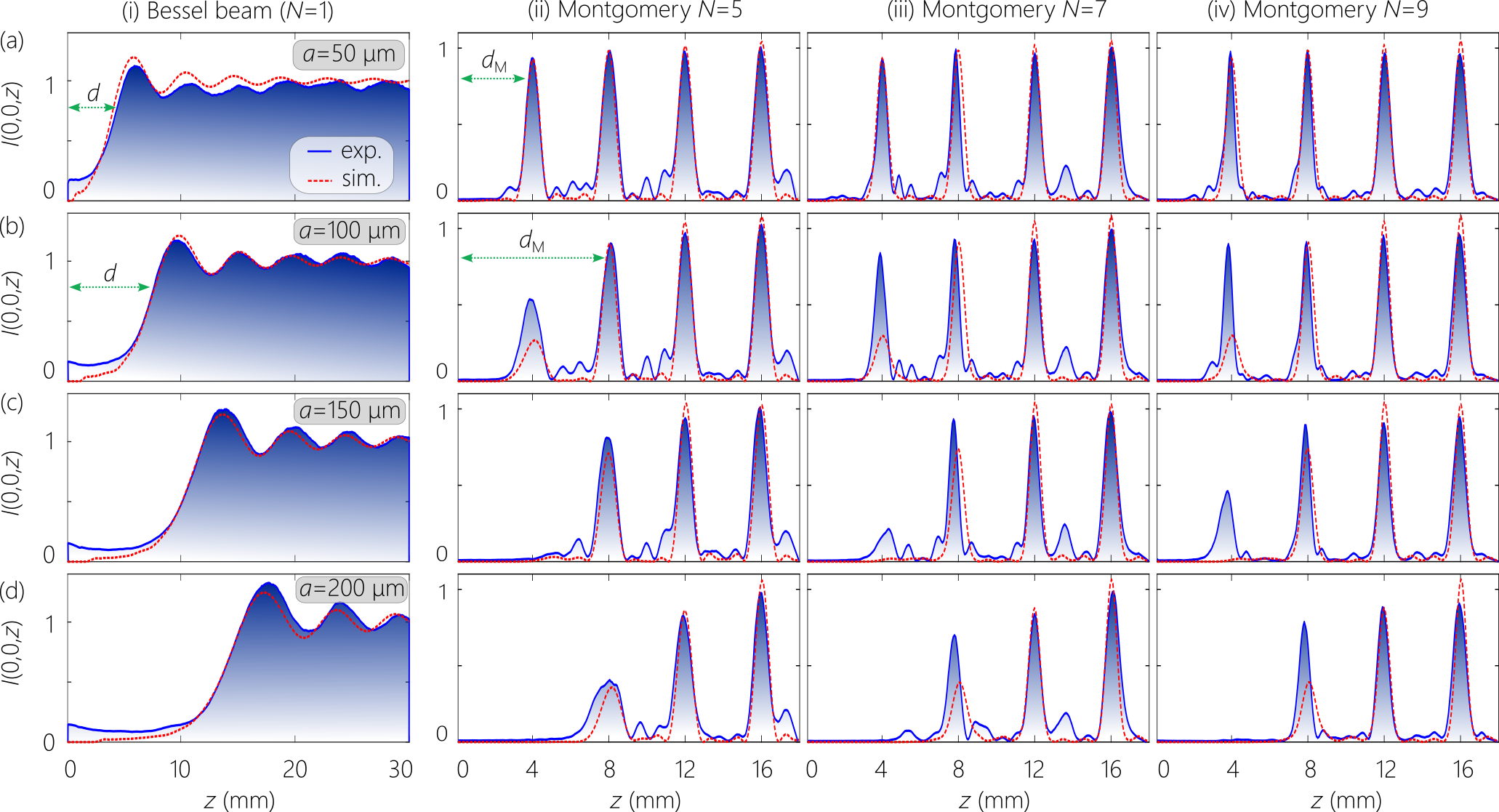}
\caption{Measured on-axis intensity profiles $I(0,0,z)$ for the Bessel beam (column i), Montgomery patterns with the self-imaging distance of $z_s\!=\!4$~mm and number of rings $N=5$ (column ii), $N=7$ (column iii), and $N=9$ (column iv). For each column, we perform the measurements for the circular obstruction radii of (a) $a\!=\!50$~\unit{\um}, (b) $a=100$~\unit{\um}, (c) $150$~\unit{\um}, and (d) $200$~\unit{\um}. The filled lines represent the experimental measurements and red dashed lines correspond to simulation plots.}
\label{Fig:Fig3}\vspace{-5mm}
\end{figure*}

\begin{equation}\label{Eq:Intensity_Montgomery}
I(r,z)=\langle|\psi(r,z)|^2\rangle=I_d(r)+I_o(r,z),
\end{equation}

here $\langle\cdot\rangle$ is the ensemble average. The diagonal contribution $I_d(r)$ is the \textit{incoherent} sum of all Bessel beams constituting the Montgomery pattern $I_d(r)=\sum_{n=1}^{N} |c_n|^2\left|\Jzero^{(n)}(r)\right|^2$, 
and is \textit{independent} of $z$. Therefore, a field with only diagonal term contributions $I(r,z)=I_d(r)$ is endowed with the diffraction-free property [Fig.~\ref{Fig:Concept} (d)]. One way to access this is via exploiting mutually incoherent rings $k_r(n)$ or assigning each ring to a different wavelength $k_r(\lambda)$ \cite{Yessenov19Optica,Yessenov22NC}. Their extended depth of focus and suppressed side-lobes make them a useful tool for fluorescent microscopy \cite{Ebrahimi2021OE,Chen2024NatMet}.  On the other hand, the off-diagonal contribution 

\begin{equation}\label{Eq:Off-DiagonalIntensity}
\begin{split}
    &I_o(r,z)\!=\\
   & \!2\!\!\sum_{n=1}^{N}\!\sum_{m>n}^{N}\langle|c_n||c_m|\Jzero^{(n)}(r)\Jzero^{(m)}(r)\cos{\left[\phi_{n,m}\!-\!\!\ks(n\!-\!m)z\right]}\rangle,
\end{split}
\end{equation}

is a cross term between different Montgomery rings, which can be observed only when they are \textit{mutually} \textit{coherent}; here $\phi_{n,m}\!=\!\arg{\{c_n\}}\!-\!\arg{\{c_m\}}$.  It is also worth noting that for a fully coherent case with $N\!>\!2$, the off-diagonal contribution $I_o$ dominates since there are $N$ diagonal terms, while $N^2\!-\!N$ off-diagonal terms in the sum; note the colormap scales in Fig.~\ref{Fig:Concept} (d,e). The relative phase between the rings $\phi_n-\phi_m$ unlocks at-will structuring of transverse profile at each plane, as it has been recently observed \cite{Yessenov26Optica}. More importantly, the off-diagonal term is responsible for the \textit{self-imaging} at discrete axial positions separated by $z_s=2\pi/k_s$ [Fig.~\ref{Fig:Concept} (e)]. Therefore, within this work, we focus on the case of mutually coherent Montgomery rings with both diagonal and off-diagonal contributions in Eq.~\ref{Eq:Intensity_Montgomery}.

Now we look into the self-healing mechanism of the Montgomery pattern. Consider a beam block placed at $z\!=\!0$ characterized by a transmission function $T(x,y)\!=\!1\!-\!\tau(x,y)$, where $\tau(x,y)$ determines the opacity function of the obstruction. According to Babinet's principle, the field profile behind the beam block can be represented by $\psi_{\B}(r,z)\!=\!\psi(r,z)T(r)\!=\!\psi(r,z)\!-\!\psi_{\A}(r,z)$, 
where $\psi_{\A}(r,z)\equiv\psi(r,z)\tau(r)$ is a complementary apertured field [Fig.~\ref{Fig:Concept} (f,g)]. From this treatment, analysis of the self-healing mechanism reduces to the diffraction profile of the complementary aperture field $\psi_{\A}(x,y)$ [Fig.~\ref{Fig:Concept} (e)]. The self-healing distance, therefore, is considered as a minimum distance $d$ at which the field passing through the obstruction recovers $\psi_{\B}(r,d)\approx \psi(r,d)$, which is equivalent to $\psi_{\A}(r,d)\approx0$. The self-healing distance $d$ is then dependent on the shape of the obstruction $\tau(r)$, the field profile $\psi(r,0)$ at $z=0$, and $\min{\{\psi_{\A}(r,d)\}}$ we consider the field to be reconstructed. For a Bessel beam $\Jzero^{(n)}(r)$ of index $n$ with propagation angle $\theta_n$, self-healing distance $d_n$ generically can be represented by  $d_{n}=\frac{\gamma}{\tan{\theta_n}}$, where the factor $\gamma$ encodes the aperture shape, field profile and recovery threshold $\min{\{\psi_{\A}(r,d)\}}$. Since the Montgomery pattern can be treated as a coherent sum of Bessel beams $c_n\Jzero^{(n)}(r)$, a heuristic approach would suggest that the self-healing distance $d_{\M}$ of the Montgomery pattern would simply be a weighted sum of $d_n$ via $d=\sum_{n=1}^{N}|c_n|^2d_{n}$. That would indeed be the case if we were to consider only the diagonal intensity term $I_d(r)$ in Eq.~\ref{Eq:Intensity_Montgomery} (incoherent sum of Bessel beams). However, the caveat is in the off-diagonal term $I_o(r,z)$ given by Eq.~\ref{Eq:Off-DiagonalIntensity} that dominates the total intensity for $N>2$. Each term in the sum would be associated with two different self-healing distances $d_n$ and $d_m$; therefore, it is not a straightforward sum of terms associated with each Bessel beam. Moreover, due to the self-imaging nature of the Montgomery pattern, the self-healing can be observed at discrete axial planes $z=z_sq$. Therefore, the self-healing distance is quantized and can only take integer multiples of $z_s$. Using the wave-optics description of the self-healing mechanism \cite{Aiello14OL}, one can derive the self-healing distance of the Montgomery pattern to take the form

\begin{equation}\label{Eq:self_healing_dis_Mont}
    d_{\M}=hz_s,\qquad h=\left\lceil\frac{\gamma}{\vartheta \zs} \right\rceil,
\end{equation}

where $\vartheta^2=2\left(\sum_{n=1}^{N} \eta_{n}\tan{\theta_n}\right)^2-\sum_{n=1}^{N} \eta_{n}\tan^2{\theta_n}$ is generalized effective propagation angle of the Montgomery pattern (the RMS angular spread weighted by spectral power) and $\eta_n\propto\frac{|\widetilde{c}_n|^2}{\sum_{n=1}^{N}|\widetilde{c}_n|^2}$ is the normalized power of each Montgomery ring, $\lceil\cdot\rceil$ is the ceiling function (see \cite{Supplementary} for derivation steps). This is one of the main findings of the paper that we experimentally confirm next. In the experiment, we use a circular disk obstruction of radius $a$, whose aperture function can be modeled by $\tau(r)=\Theta(a-r)$; $\Theta(x)$ is the Heaviside step-function. 


The experimental setup is shown in Fig.~\ref{Fig:Fig2}(a). A monochromatic laser source at $\lambda = 532~\mathrm{nm}$ is expanded to a diameter of approximately $20~\mathrm{mm}$. The expanded beam illuminates a reflective spatial light modulator (SLM; Santec 200), which encodes the axicon-like phase required to generate a Montgomery effect [Fig.~\ref{Fig:Fig2}(a)]. The modulated beam is retro-reflected and directed through a 50:50 beam splitter (BS), which sends the retro-reflected beam from the SLM into the characterization path. 
The beam passes through a $2.5\times$ de-magnification system with spherical lenses L$_1$ and L$_2$ of focal lengths $f_1=250~\mathrm{mm}$ and $f_2=100~\mathrm{mm}$, respectively, and is observed on a camera (CCD, TheImagingSource DMK 37AUX226) with an objective lens attached for improved resolution. We place a spatial filter at the Fourier plane of the lens L$_1$, blocking the zeroth order to filter out unmodulated intensity from the SLM. We scan the camera along $z$ with a motorized translation stage (Thorlabs LTS150), recording the transverse intensity profile $I(x,y)$ at each $z$ [Fig.~\ref{Fig:Fig2}(b)], from which we reconstruct the axial intensity profile $I(x,y,z)$ [Fig.~\ref{Fig:Fig2}(c)]. To observe the self-healing behind a beam block, we place an opaque circular disk of radius $a$ at $z=0$ and scan the intensity profiles behind it [Fig.~\ref{Fig:Fig2}(d-h)]. We use Thorlabs full aperture obstruction targets of various radii $a\!=\!50\!-\!250$~\unit{\um} (Thorlabs, R1D100P-R1D500P) as a beam block.

\begin{figure}[t!]
\centering
\includegraphics[width=8.6cm]{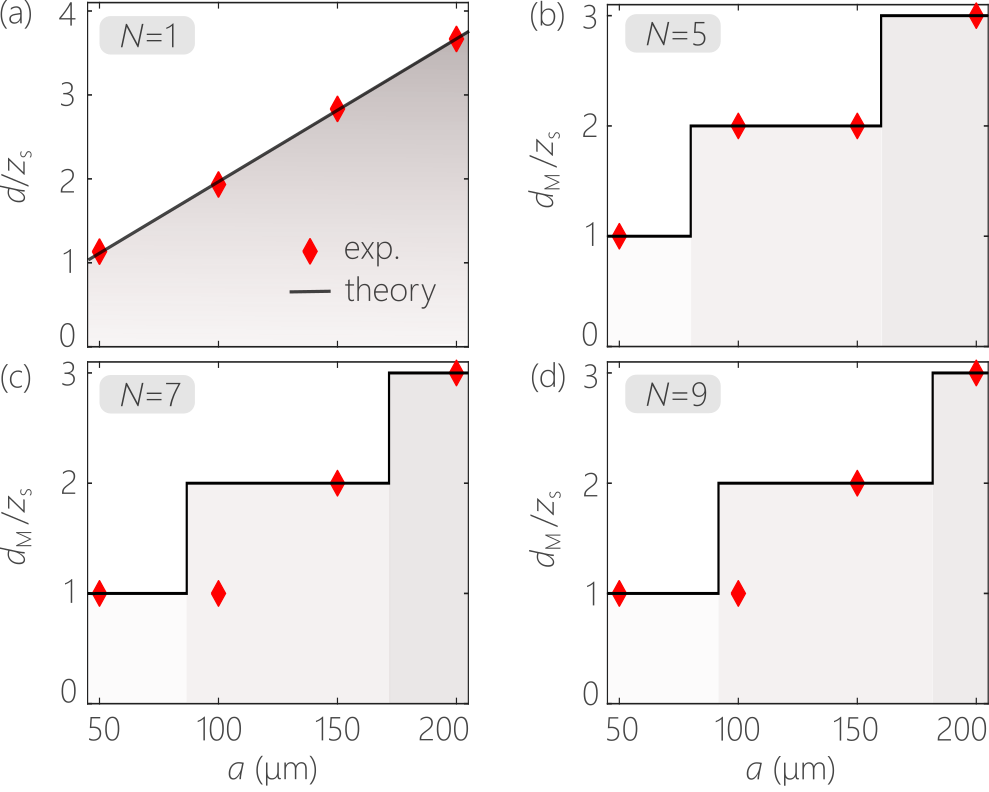}
\caption{Self-healing distances retrieved from the on-axis measurements in Fig.~\ref{Fig:Fig3} for (a) Bessel beam and Montgomery patterns of (b) $N\!=\!5$, (c) $N\!=\!7$, and (d) $N\!=\!9$. We consider the field to be recovered if the normalized intensity after the block reaches $80\%$ of its initial no-block intensity, i.e., $I_{\B}\!=\!0.8I$. For the Bessel beam in (a), the self-healing $d$ is continuous, whereas for the Montgomery field in (b-d), the self-healing distance is at integer multiples of $z_s$: $d_{\M}\!=\!h\zs$. Red rhombi correspond to the experimental measurements, whereas black lines are theoretical expectations from Eq.~\ref{Eq:self_healing_dis_Mont}.}\vspace{-5mm}
\label{Fig:Fig4}
\end{figure}

To demonstrate the self-healing property we first synthesize a Montgomery pattern with $N\!=\!7$ rings and the self-imaging distance of $\zs\!=\!10$~mm [Fig.~\ref{Fig:Fig2}(b,c)], and observe four revivals of a spot with a beam radius of $\Delta r\!\approx\!18$~\unit{\um} ($1/e^2$ radius) at $z\!=\!0,10,20,30$~mm. Then we place a circular disk block of radius $a$ at the beam center $(x,y,z)\!=\!(0,0,0)$ [Fig.~\ref{Fig:Fig2}(d-h)]. For the block size of $a\!=\!50$~\unit{\um}$\approx\!2.8\!\times\!\Delta r$ [Fig.~\ref{Fig:Fig2}(d)] and $a\!=\!100$~\unit{\um} [Fig.~\ref{Fig:Fig2}(e)], the first peak at $z\!=\!\zs$ fully recovers. When the obstruction size is further increased to $a\!=\!150$~\unit{\um} [Fig.~\ref{Fig:Fig2}(f)] and $a\!=\!200$~\unit{\um} [Fig.~\ref{Fig:Fig2}(g)], the first peak starts to deteriorate, while the second peak at $z\!=\!2\zs\!=\!\!20$~mm stays intact. Finally, when the block size is $a\!=\!250$~\unit{\um}$\approx\!14\!\times\!\Delta r$, the first peak completely disappears, while the second peak is mostly undamaged [Fig.~\ref{Fig:Fig2}(h)].

We repeat the measurements in Fig.~\ref{Fig:Fig2}(d-h) for the Bessel beam [Fig.~\ref{Fig:Fig3}, column (i)], which is synthesized by the same setup via isolating only the first Montgomery ring in Fig.~\ref{Fig:Concept}(a). For the quantitative analysis of the self-healing distance $d_{\M}$, in Fig.~\ref{Fig:Fig3} we plot the on-axis intensity profiles $I(z)\!=\!I(0,0,z)$, where the filled blue line represents experimental measurements and the red dashed line corresponds to simulation results. We define the self-healing distance $d$ as the minimum distance required for the blocked beam to recover to $80\%$ of its initial intensity (green dotted arrows in Fig.~\ref{Fig:Fig3} (a,b)). The Bessel beam displays a characteristic ripple-like profile after it recovers \cite{chu_analytical_2012}, with the self-healing distance increasing from $d\!\approx\!4.5$~mm to $d\!\approx\!15$~mm as the block size is increased from $a\!=\!50$~\unit{\um} to $a\!=\!200$~\unit{\um}. Same measurements are repeated for the Montgomery patterns of self-imaging distance $\zs\!=\!4$~mm with $N\!=\!5$ [Fig.~\ref{Fig:Fig3}, column (ii)], $N\!=\!7$ [Fig.~\ref{Fig:Fig3}, column (iii)] and $N\!=\!9$ [Fig.~\ref{Fig:Fig3}, column (iv)], with the corresponding beam sizes of $\Delta r\!\approx\!13$~\unit{\um}, $\Delta r\!\approx\!11$~\unit{\um} and $\Delta r\!\approx\!10$~\unit{\um}, respectively. For the block size of $a\!=\!50$~\unit{\um}, all of the Montgomery patterns remain practically unaffected [Fig.~\ref{Fig:Fig3} (a)]. For the block size of $a\!=\!100$~\unit{\um}, the first self-imaging plane at $z\!=\!4$~mm of the Montgomery pattern with $N\!=\!5$ does not recover [Fig.~\ref{Fig:Fig3} (b), column (ii)], therefore the self-healing distance is $d_{\M}\!=\!4$~mm.  These measurements highlight the main difference in the self-healing mechanism of the Montgomery pattern, as the self-healing plane coincides with one of the self-imaging planes. Finally, we plot the measurements of self-healing distances $d$ and $d_{\M}$ in Fig.~\ref{Fig:Fig4} and compare with the theoretical expectations. For the Bessel beam [Fig.~\ref{Fig:Fig4} (a)], the self-healing distance fits well with the theoretical expectations for $\gamma\!\approx\!1.1$, which we use for the rest of Fig.~\ref{Fig:Fig4}. Measurements of $d_{\M}$ for the Montgomery patterns, which take integer multiple values of $z_s$, match well with the theoretical expectations given by Eq.~\ref{Eq:self_healing_dis_Mont}, except for a few outliers that we anticipate to be caused by alignment imperfections and aberrations in the system.  

In conclusion, we have demonstrated theoretically and experimentally that the Montgomery self-imaging pattern exhibits self-healing upon traversal of an opaque circular obstruction. Using Babinet's principle, we analyzed the self-healing of the Montgomery field as a coherent superposition of underlying Bessel beams with square-root spatial frequency distribution. We derived a closed-form expression for the self-healing distance, which depends on the obstruction size, number of rings $N$, and is quantized with the self-imaging distance $z_s$. This quantization is a qualitative distinction from the continuous self-healing of diffraction-free beams, and is a direct consequence of the discrete axial structure intrinsic to self-imaging fields. Our measurements confirm the predicted quantization across a range of obstruction radii $a$ and ring numbers $N$. These results establish the resilience of the Montgomery pattern against scatterers and obstructions, strengthening its prospect as a platform for multi-plane optical manipulation~\cite{garces-chavez_simultaneous_2002}, light-sheet imaging in scattering media~\cite{Fahrbach10NP}, and plane-selective control of neutral-atom arrays~\cite{Schlosser2023PRL,chiu2025Nature}, and compact beam-shaping and point-singularity array implementations via flat optics \cite{Dorrah2022Science, Dorrah2025NC,lim_point_2023}.

\textbf{Funding}
Work at Harvard was supported by the ONR MURI program, under award N00014-20-1-2450, and by the Air Force Office of Scientific Research (AFOSR) under award FA9550-22-1-0243. Work at UCF was supported by the US Office of Naval Research (ONR) under award N00014-17-1-2458, and the ONR MURI program under award N00014-20-1-2789.  

\textbf{Disclosures}
The authors declare no conflicts of interest.

\textbf{Data availability}
Data underlying the results presented in this paper are not publicly available at this time but may be obtained from the authors upon reasonable request.


\bibliography{Montgomery}

@ARTICLE{Durnin87PRL,
  AUTHOR =       {J. Durnin and J. J. Miceli and J. H. Eberly},
  TITLE =        {Diffraction-free beams},
  JOURNAL =      {Phys. Rev. Lett.},
  YEAR =         {1987},
  volume =       {58},
  pages =        {1499-1501},
}

@ARTICLE{Kondakci17NP,
  AUTHOR =       {H. E. Kondakci and A. F. Abouraddy},
  TITLE =        {Diffraction-free space-time beams},
  JOURNAL =      {Nat. Photon.},
  YEAR =         {2017},
  volume =       {11},
  pages =        {733-740},
}

@ARTICLE{Yessenov22AOP,
  AUTHOR =       {M. Yessenov and L. A. Hall and K. L. Schepler and A. F. Abouraddy},
journal = {Adv. Opt. Photon.},
number = {3},
pages = {455--570},
publisher = {Optica Publishing Group},
title = {Space-time wave packets},
volume = {14},
month = {Sep},
year = {2022},
url = {https://opg.optica.org/aop/abstract.cfm?URI=aop-14-3-455},
doi = {10.1364/AOP.450016},
}

@article{Yessenov22NC,
  title={Space-time wave packets localized in all dimensions},
  author={M. Yessenov and J. Free and Z. Chen and E. G. Johnson and M. P. J. Lavery and M. A. Alonso and A. F. Abouraddy},
  journal={Nat. Commun.},
  volume={13},
  pages={4573},
  year={2022},
}

@ARTICLE{Yessenov19Optica,
  AUTHOR =       {M. Yessenov and B. Bhaduri and H. E. Kondakci and M. Meem and R. Menon and A. F. Abouraddy},
  TITLE =        {Non-diffracting broadband incoherent space-time fields},
  JOURNAL =      {Optica},
  YEAR =         {2019},
  volume =       {6},
  pages =        {598-607},
}

@ARTICLE{Kondakci18OL,
  AUTHOR =       {H. E. Kondakci and A. F. Abouraddy},
  TITLE =        {Self-healing of space-time light sheets},
  JOURNAL =      {Opt. Lett.},
  YEAR =         {2018},
  volume =       {43},
  pages =        {3830-3833},
}

@ARTICLE{Yessenov19PRA,
  AUTHOR =       {M. Yessenov and B. Bhaduri and H. E. Kondakci and A. F. Abouraddy},
  TITLE =        {Classification of propagation-invariant space-time light-sheets in free space: Theory and experiments},
  JOURNAL =      {Phys. Rev. A},
  YEAR =         {2019},
  volume =       {99},
  pages =        {023856},

}

@ARTICLE{Siviloglou07OL,
  AUTHOR =       {G. A. Siviloglou and D. N. Christodoulides},
  TITLE =        {Accelerating finite energy {A}iry beams},
  JOURNAL =      {Opt. Lett.},
  YEAR =         {2007},
  volume =       {32},
  pages =        {979-981},
}

@ARTICLE{Fahrbach10NP,
  AUTHOR =       {F. O. Fahrbach and P. Simon and A. Rohrbach},
  TITLE =        {Microscopy with self-reconstructing beams},
  JOURNAL =      {Nat. Photon.},
  YEAR =         {2010},
  volume =       {4},
  pages =        {780-785},
}

@ARTICLE{Broky08OE,
  AUTHOR =       {J. Broky and G. A. Siviloglou and A. Dogariu and D. N. Christodoulides},
  TITLE =        {Self-healing properties of optical {A}iry beams},
  JOURNAL =      {Opt. Express},
  YEAR =         {2008},
  volume =       {16},
  pages =        {12880-12891},
}

@article{lim_point_2023,
	title = {Point singularity array with metasurfaces},
	volume = {14},
	issn = {2041-1723},
	url = {https://www.nature.com/articles/s41467-023-39072-6},
	doi = {10.1038/s41467-023-39072-6},
	pages = {3237},
	number = {1},
	journal = {Nat. Comm.},
	shortjournal = {Nat Commun},
	author = {Lim, Soon Wei Daniel and Park, Joon-Suh and Kazakov, Dmitry and Spägele, Christina M. and Dorrah, Ahmed H. and Meretska, Maryna L. and Capasso, Federico},
	urldate = {2025-04-14},
	date = {2023-06-05},
	langid = {english},
}

@article{lohmann_fractional_2005,
	title = {Fractional {M}ontgomery effect: a self-imaging phenomenon},
	volume = {22},
	rights = {© 2005 Optical Society of America},
	issn = {1520-8532},
	url = {https://opg.optica.org/josaa/abstract.cfm?uri=josaa-22-8-1500},
	doi = {10.1364/JOSAA.22.001500},
	shorttitle = {Fractional Montgomery effect},
	pages = {1500--1508},
	number = {8},
	journal = {{JOSA} A},
	shortjournal = {J. Opt. Soc. Am. A, {JOSAA}},
	author = {Lohmann, Adolf W. and Knuppertz, Hans and Jahns, Jürgen},
	urldate = {2025-05-20},
	date = {2005-08-01},
    year ={2005},
}

@article{jahns_montgomery_2003,
	title = {Montgomery self-imaging effect using computer-generated diffractive optical elements},
	volume = {225},
	issn = {0030-4018},
	url = {https://www.sciencedirect.com/science/article/pii/S0030401803018224},
	doi = {10.1016/j.optcom.2003.07.033},
	pages = {13--17},
	number = {1},
	journal = {Optics Communications},
	shortjournal = {Optics Communications},
	author = {Jahns, Jürgen and Knuppertz, Hans and Lohmann, Adolf W.},
	urldate = {2025-05-20},
	date = {2003-09-15},
}

@ARTICLE{Talbot36PM,
  AUTHOR =       {H. F. Talbot},
  TITLE =        {Facts relating to optical science. {N}o. {IV}},
  JOURNAL =      {Philos. Mag.},
  YEAR =         {1836},
  volume =       {9},
  pages =        {401-407},
}

@ARTICLE{Montgomery67JOSA,
  AUTHOR =       {W. D. Montgomery},
  TITLE =        {Self-imaging objects of infinite aperture},
  JOURNAL =      {J. Opt. Soc. Am.},
  YEAR =         {1967},
  volume =       {57},
  pages =        {772-778},
}

@article{Shen2023Roadmap,
  title={Roadmap on spatiotemporal light fields},
  author={Shen, Yijie and Zhan, Qiwen and Wright, Logan G and Christodoulides, Demetrios N and Wise, Frank W and Willner, Alan E and Zou, Kai-heng and Zhao, Zhe and Porras, Miguel A and Chong, Andy and others},
  journal={Journal of Optics},
  volume={25},
  number={9},
  pages={093001},
  year={2023},
  publisher={IOP Publishing}
}

@article{Schlosser2023PRL,
  title={Scalable multilayer architecture of assembled single-atom qubit arrays in a three-dimensional Talbot tweezer lattice},
  author={Schlosser, Malte and Tichelmann, Sascha and Sch{\"a}ffner, Dominik and de Mello, Daniel Ohl and Hambach, Moritz and Sch{\"u}tz, Jan and Birkl, Gerhard},
  journal={Physical Review Letters},
  volume={130},
  number={18},
  pages={180601},
  year={2023},
  publisher={APS}
}

@article{De2021JoP,
  title={A versatile ring trap for quantum gases},
  author={de Herve, Mathieu de Go{\"e}r and Guo, Yanliang and De Rossi, Camilla and Kumar, Avinash and Badr, Thomas and Dubessy, Romain and Longchambon, Laurent and Perrin, H{\'e}l{\`e}ne},
  journal={Journal of Physics B: Atomic, Molecular and Optical Physics},
  volume={54},
  number={12},
  pages={125302},
  year={2021},
  publisher={IOP Publishing}
}

@article{Kusano2025PhysRevRes,
  title={Plane-selective manipulations of nuclear spin qubits in a three-dimensional optical tweezer array},
  author={Kusano, Toshi and Nakamura, Yuma and Yokoyama, Rei and Ozawa, Naoya and Shibata, Kosuke and Takano, Tetsushi and Takasu, Yosuke and Takahashi, Yoshiro},
  journal={Physical Review Research},
  volume={7},
  number={2},
  pages={L022045},
  year={2025},
  publisher={APS}
}

@article{Dorrah2022Science,
  title={Tunable structured light with flat optics},
  author={Dorrah, Ahmed H and Capasso, Federico},
  journal={Science},
  volume={376},
  number={6591},
  pages={eabi6860},
  year={2022},
  publisher={American Association for the Advancement of Science}
}

@article{Dorrah2025NC,
  title={Free-standing bilayer metasurfaces in the visible},
  author={Dorrah, Ahmed H and Park, Joon-Suh and Palmieri, Alfonso and Capasso, Federico},
  journal={Nat. Comm.},
  volume={16},
  number={1},
  pages={3126},
  year={2025},
  publisher={Nature Publishing Group UK London}
}

@article{Ryu13PRL,
  title={Experimental Realization of {J}osephson {J}unctions for an Atom {SQUID}},
  author={C. Ryu and P. W. Blackburn and A. A. Blinova and M. G. Boshier},
  journal={Phys. Rev. Lett.},
  volume={111},
  pages={205301},
  year={2013},
}

@article{Indebetouw1992JOSAA,
  title={Quasi-self-imaging using aperiodic sequences},
  author={Indebetouw, Guy},
  journal={Journal of the Optical Society of America A},
  volume={9},
  number={4},
  pages={549--558},
  year={1992},
  publisher={Optical Society of America}
}

@ENTRY{Supplementary,
  title =       {Supplemental Material [url]}
}

@article{Chen2024NatMet,
  title={High-throughput volumetric mapping of synaptic transmission},
  author={Chen, Wei and Ge, Xinxin and Zhang, Qinrong and Natan, Ryan G and Fan, Jiang Lan and Scanziani, Massimo and Ji, Na},
  journal={Nature Methods},
  volume={21},
  number={7},
  pages={1298--1305},
  year={2024},
  publisher={Nature Publishing Group US New York}
}

@article{chiu2025Nature,
  title={Continuous operation of a coherent 3,000-qubit system},
  author={Chiu, Neng-Chun and Trapp, Elias C and Guo, Jinen and Abobeih, Mohamed H and Stewart, Luke M and Hollerith, Simon and Stroganov, Pavel L and Kalinowski, Marcin and Geim, Alexandra A and Evered, Simon J and others},
  journal={Nature},
  pages={1--3},
  year={2025},
  publisher={Nature Publishing Group UK London}
}

@article{Aiello14OL,
author = {Andrea Aiello and Girish S. Agarwal},
journal = {Opt. Lett.},
number = {24},
pages = {6819--6822},
publisher = {Optica Publishing Group},
title = {Wave-optics description of self-healing mechanism in {B}essel beams},
volume = {39},
month = {Dec},
year = {2014},
url = {https://opg.optica.org/ol/abstract.cfm?URI=ol-39-24-6819},
doi = {10.1364/OL.39.006819},
}

@article{Yessenov26Optica,
author = {Murat Yessenov and Luca Sacchi and Alfonso Palmieri and Layton A. Hall and Ayman F. Abouraddy and Federico Capasso},
journal = {Optica},
number = {2},
pages = {195--202},
publisher = {Optica Publishing Group},
title = {Observation of the spatially structured {M}ontgomery effect in free space},
volume = {13},
month = {Feb},
year = {2026},
url = {https://opg.optica.org/optica/abstract.cfm?URI=optica-13-2-195},
doi = {10.1364/OPTICA.582198},
}

@article{Ebrahimi2021OE,
  title={Incoherent superposition of polychromatic light enables single-shot nondiffracting light-sheet microscopy},
  author={Ebrahimi, Vahid and Tang, Jialei and Han, Kyu Young},
  journal={Optics Express},
  volume={29},
  number={20},
  pages={32691--32699},
  year={2021},
  publisher={Optical Society of America}
}

@article{chu_analytical_2012,
    title = {Analytical study on the self-healing property of {Bessel} beam},
    volume = {66},
    issn = {1434-6079},
    url = {https://doi.org/10.1140/epjd/e2012-30343-6},
    doi = {10.1140/epjd/e2012-30343-6},
    language = {en},
    number = {10},
    urldate = {2026-04-17},
    journal = {The European Physical Journal D},
    author = {Chu, X.},
    month = oct,
    year = {2012},
    pages = {259},
}

@article{garces-chavez_simultaneous_2002,
    title = {Simultaneous micromanipulation in multiple planes using a self-reconstructing light beam},
    volume = {419},
    copyright = {2002 Macmillan Magazines Ltd.},
    issn = {1476-4687},
    url = {https://www.nature.com/articles/nature01007},
    doi = {10.1038/nature01007},
    language = {en},
    number = {6903},
    urldate = {2026-04-17},
    journal = {Nature},
    publisher = {Nature Publishing Group},
    author = {Garcés-Chávez, V. and McGloin, D. and Melville, H. and Sibbett, W. and Dholakia, K.},
    month = sep,
    year = {2002},
    pages = {145--147},
}

@article{baliyan_unveiling_2025,
    title = {Unveiling the self-healing potential of {Bessel}–{Gauss} beams in image encryption},
    volume = {42},
    copyright = {\&\#169; 2024 Optica Publishing Group},
    issn = {1520-8532},
    url = {https://opg.optica.org/josaa/abstract.cfm?uri=josaa-42-1-36},
    doi = {10.1364/JOSAA.544736},
    language = {EN},
    number = {1},
    urldate = {2026-04-17},
    journal = {JOSA A},
    publisher = {Optica Publishing Group},
    author = {Baliyan, Mansi and Nishchal, Naveen K.},
    month = jan,
    year = {2025},
    pages = {36--44},
}

@article{Bouchal1998OC,
  title={Self-reconstruction of a distorted nondiffracting beam},
  author={Bouchal, Zdenek and Wagner, J and Chlup, M},
  journal={Optics Communications},
  volume={151},
  number={4-6},
  pages={207--211},
  year={1998},
  publisher={Elsevier}
}

\end{document}


\title{Supplementary material for ``Self-healing of the Montgomery pattern''}

\author{Athena Xu}
\affiliation{Harvard John A. Paulson School of Engineering and Applied Sciences, Harvard University, Cambridge, MA, USA}
\affiliation{Department of Electrical and Computer Engineering, University of Waterloo, Waterloo, N2L3G1, ON, Canada}

\author{Oscar de Vries} 
\affiliation{Harvard John A. Paulson School of Engineering and Applied Sciences, Harvard University, Cambridge, MA, USA}
\affiliation{Department of Applied Physics and Science Education, Eindhoven University of Technology, 5600 MB Eindhoven, The Netherlands}

\author{Alfonso Palmieri}
\affiliation{Harvard John A. Paulson School of Engineering and Applied Sciences, Harvard University, Cambridge, MA, USA}

\author{Murat Yessenov}
\email{yessenov@seas.harvard.edu}
\affiliation{Harvard John A. Paulson School of Engineering and Applied Sciences, Harvard University, Cambridge, MA, USA}
\affiliation{CREOL, The College of Optics \& Photonics, University of Central Florida, Orlando, Florida 32816, USA}

\author{Ayman F. Abouraddy}
\affiliation{CREOL, The College of Optics \& Photonics, University of Central Florida, Orlando, Florida 32816, USA}

\author{Federico Capasso}
\email{capasso@seas.harvard.edu}
\affiliation{Harvard John A. Paulson School of Engineering and Applied Sciences, Harvard University, Cambridge, MA, USA}

\date{\today}
\renewcommand{\thepage}{S\arabic{page}}  
\renewcommand{\thefigure}{S\arabic{figure}}
\renewcommand{\theequation}{S\arabic{equation}}
\renewcommand{\thetable}{Supplementary Table~\arabic{table}}

\maketitle

\tableofcontents

\clearpage

\section{Theory of the Montgomery effect}

Consider a monochromatic scalar optical field of angular frequency $\omega_o$ propagating in free space along the $z$ axis. In cylindrical coordinates $(r,\phi,z)$,
\begin{equation}
E(r,\phi,z;t)=\psi(r,\phi,z)\e^{i(k_o z-\omega_o t)},
\end{equation}
where $k_o=\omega_o/c$ with $c$ being the speed of light in a vacuum, and the slowly varying envelope is written in angular-spectrum form as
\begin{equation}
\psi(r,\phi,z)=\int_0^{\infty} \dd \kr \int_0^{2\pi} \dd \chi\; \kr\, \widetilde{\psi}(\kr,\chi)
\e^{i \kr r \cos(\phi-\chi)}\e^{-i(k_o-\kz)z}.
\end{equation}
Here $(\kr,\chi,\kz)$ are cylindrical spectral coordinates, with $\kr=\sqrt{k_x^2+k_y^2}$ and $\chi=\arctan(k_y/k_x)$; $\widetilde{\psi}(k_{r},\chi)$ is the spatial spectrum of the beam. Because the angular–spectrum integral features an integrand that is periodic in the spectral angle \(\chi\) with period \(2\pi\), we may expand the spectral amplitude
\(\widetilde{\psi}(k_{r},\chi)\) in a Fourier series in \(\chi\).  Specifically, we can write
\begin{equation}
\widetilde{\psi}(\kr,\chi)=\frac{1}{2\pi}\sum_{\ell=-\infty}^{\infty}\widetilde{\psi}_{\ell}(\kr)\e^{i \ell \chi},
\end{equation}
where each coefficient \(\widetilde{\psi}_{\ell}(k_{r})\)
is the radial spectrum of the \(\ell\)th azimuthal harmonic and the normalization \(1/(2\pi)\) ensures that the inverse relation is
 \(\widetilde{\psi}_{\ell}(k_{r})
=\int_{0}^{2\pi}\widetilde{\psi}(k_{r},\chi)\,e^{-i\ell\chi}\,\mathrm d\chi\).
Here we focus on rotationally symmetric fields with $\ell=0$, therefore $\widetilde{\psi}(\kr,\chi)=\widetilde{\psi}(\kr)$ and
\begin{equation}
\psi(r,z)=\int_0^{\infty} \dd \kr\; \kr\,\widetilde{\psi}(\kr)\Jzero(\kr r)\e^{-i(k_o-\kz)z},
\end{equation}
where $J_{0}$ is the zeroth-order Bessel function of the first kind. 

Self-imaging at planes $z=q\zs$ with integer $q$ requires
\begin{equation}
(k_o-\kz)\zs=2\pi n,
\qquad n=1,2,\dots,N;
\end{equation}
where $N=\bigl\lfloor k_{o}/k_{s}\bigr\rfloor$ is the upper-bound of the sampling range.
In the paraxial regime,
\begin{equation}
\kz\approx k_o-\frac{\kr^2}{2k_o},
\end{equation}
so the Montgomery sampling law is
\begin{equation}\label{Eq:MontCondition}
\kz(n)=k_o-\ks n,
\qquad
\kr(n)=k_{\mathrm{L}}\sqrt{n},
\end{equation}
with
\begin{equation}
\ks=\frac{2\pi}{\zs},
\qquad
k_{\mathrm{L}}=\sqrt{2k_o\ks}=2\pi\sqrt{\frac{2}{\zs\lambda}}.
\end{equation}
Therefore, the spatial spectrum of an ideal Montgomery field is
\begin{equation}\label{Eq:Montg_spectrum}
\widetilde{\psi}_{\mathrm{M}}(\kr)=\sum_{n=1}^{N} \widetilde{c}_n\,\delta\!\bigl(\kr-k_{\mathrm{L}}\sqrt{n}\bigr),
\end{equation}
which corresponds to the concentric rings in the spectral space, known as Montgomery rings; here $\widetilde{c}_n$ is the complex coefficient of each Montgomery ring. The corresponding field becomes the coherent superposition of Bessel beams
\begin{equation}\label{Eq:Montgomery_field}
\psi_{\mathrm{M}}(r,z)=\sum_{n=1}^{N} c_n\Jzero\!\bigl(k_{\mathrm{L}}\sqrt{n}\,r\bigr)\e^{-i \ks n z},
\end{equation}
where each Montgomery ring $k_r(n)$ generates a Bessel beam $J_{0}(k_r(n)\,r)$ in the physical space, and $c_n=\widetilde{c}_nk_r(n)$ is the complex weight of each Bessel beam that can be calculated from Parseval's theorem. 
For convenience, we can represent Eq.~\ref{Eq:Montgomery_field} in the form of vector multiplication via
\begin{equation}
    \psi_{\mathrm{M}}(r,z)=\mathbf{j}(r)^{T}\mathbf{a}(z),
\end{equation}
where
\begin{equation}
    \mathbf{a}(z)=
\begin{bmatrix}
c_1 e^{-ik_s z}\\
\vdots\\
c_N e^{-ik_sN z}
\end{bmatrix}
,\qquad
\mathbf{j}(r)=
\begin{bmatrix}
J_0(\kL r)\\
\vdots\\
J_0(k_{\mathrm{L}}\sqrt{N} r)
\end{bmatrix}.
\end{equation}

Then the intensity profile of the Montgomery field can be represented by 
\begin{equation}\label{Eq:Intensity_Montgomery}
I_{\mathrm{M}}(r,z)=\langle|\psi_{\mathrm{M}}(r,z)|^2\rangle =\langle\mathbf{a}(z)^{\dagger}K(r)\mathbf{a}(z)\rangle. 
\end{equation}
 Here $\langle\cdot\rangle$ ensemble average, which is typically replaced with averaging in time. $K(r)=\mathbf{j}(r)\mathbf{j}(r)^{T}$ is a point-wise Gram kernel (rank-1, symmetric and real):
 \begin{equation}\label{Eq:K_r}
 K(r)=
     \begin{bmatrix}
      \Jzero^2(\kL r)  & \hdots   & \Jzero(\kL r)\Jzero(\sqrt{N}\kL r)\\
      \vdots         & \vdots   & \vdots \\
      \Jzero(\sqrt{N}\kL r)\Jzero(\kL r) & \hdots & \Jzero^2(\sqrt{N}\kL r)
      
     \end{bmatrix}
 \end{equation}
 One can note that the $z$-dependence of $\mathbf{a}(z)$ is in the phase term, and the inner product $\mathbf{a}(z)^{\dagger}\mathbf{a}(z)=\sum^{N}_{n=1}|c_n|^2$ is independent of $z$. To exploit this property, we separate contributions to the intensity from diagonal $K_{m=n}(r)$ and off-diagonal $K_{m\neq n}(r)$ elements of $K(r)$. We separate diagonal and off-diagonal elements into two separate matrices $K(r)=D(r)+O(r)$, where $D(r)=\mathrm{diag}\{K(r)\}$ and $O(r)=K(r)-D(r)$. We can then represent the intensity of the Montgomery field as a sum of contributions from diagonal and off-diagonal elements
 \begin{equation}
     I_{\M}(r,z)=I_{d}(r)+I_o(r,z),
 \end{equation}
 where 
 \begin{equation}
     \begin{split}
         I_d(r)=&\langle\mathbf{a}(z)^{\dagger}D(r)\mathbf{a}(z)\rangle=\sum_{n=1}^{N} |c_n|^2\left|\Jzero\!\bigl(k_{\mathrm{L}}\sqrt{n}\,r\bigr)\right|^2, \\
         I_o(r,z)=&\langle\mathbf{a}(z)^{\dagger}O(r)\mathbf{a}(z)\rangle=\sum_{n=1}^{N}\sum_{m\neq n}^{N} \langle c_nc^{*}_m\Jzero\!\bigl(\kL\sqrt{n}\,r\bigr)\Jzero\!\bigl(\kL\sqrt{m}\,r\bigr)\e^{-i \ks (n-m) z}\rangle=\\
         =&2\sum_{n=1}^{N}\sum_{m>n}^{N}\langle|c_n||c_m|\Jzero\!\bigl(\kL\sqrt{n}r)\Jzero\!\bigl(\kL\sqrt{m}r)\cos{\left[(\phi_n-\phi_m)-\ks(n-m)z\right]}\rangle,
     \end{split}
 \end{equation}
 and  $\phi_n=\arg{\{c_n\}}$ and $\phi_m=\arg{\{c_m\}}$. 
These two parts have quite different characteristic features that are worth discussing in more detail. The diagonal term $I_d(r)$ is the \textit{incoherent} sum of all Bessel beams constituting the Montgomery field, and it is \textit{independent} of $z$. Therefore, a field with only diagonal term contributions $I(r,z)=I_d(r)$ is endowed with the diffraction-free property. One way to achieve this is via exploiting mutually incoherent rings $k_r(n)$ or assigning each ring to a different wavelength $k_r(\lambda)$ \cite{Yessenov19Optica}. Their extended depth of focus and suppressed side-lobes make them a useful tool for fluorescent microscopy \cite{Ebrahimi2021OE}[Naji]. On the other hand, the off-diagonal contribution $I_o(r,z)$ is a cross term between different Montgomery rings, which can be observed only when they are \textit{mutually} \textit{coherent}. It is also sensitive to the relative phase between the rings $\phi_n-\phi_m$, enabling additional control over the intensity profile.   

\section{Self-healing of a Bessel beam}

Since Bessel beams are building blocks of the Montgomery field, we first consider the self-healing mechanism of a Bessel beam $J_{0}(k_r(n)\,r)$ corresponding to the $n^{\mathrm{th}}$ ring in the spectral space $k_{r}(n)$ given by Eq.~\ref{Eq:MontCondition}. For a circularly symmetric obstruction of radius $a$, a minimum reconstruction distance can be calculated via a simple ray-tracing model
\begin{equation}
 d_{n}=a\cot{\theta_{n}},
\label{Eq:z_min_Bessel}
\end{equation}
where $\theta_{n}=\arctan{[k_{r}(n)/k_{z}(n)]}$ is the propagation angle of plane-wave components of the Bessel beam. Although the expression in Eq.~\ref{Eq:z_min_Bessel} gives an estimate, it does not take into account the shape of the obstruction, therefore does not accurately determine the self-healing distance. More importantly, as the Montgomery effect is an inherently wave phenomenon, a ray optics description is not a suitable platform for its treatment. Therefore, we follow the wave-description approach presented in ref. \cite{Aiello14OL}, and extend their analysis for the Montgomery field. 

Consider an obstruction placed at $z=0$ characterized by a transmission function $T(x,y)=1-\tau_A(x,y)$, where $\tau_A(x,y)$ determines the opacity function of the obstruction. $A\in\{\square,G,\circ\}$ labels different aperture shapes that we study below: $\square$ - square , $G$ - soft-Gaussian, $\circ$ - circular disk. 

According to Babinet principle, the field profile behind the obstruction can be represented as 
\begin{equation}
 \psi_{\B}(x,y)=\psi(x,y)T(x,y)=\psi(x,y)-\psi(x,y)\tau_A(x,y)=\psi(x,y)-\psi_{A}(x,y),
\label{Eq:Bessel_obstructed}
\end{equation}
where $\psi_{A}(x,y)\equiv\psi(x,y)\tau_A(x,y)$ is a complementary apertured field. From this treatment, the analysis of the self-healing mechanism of any light field reduces to the diffraction profile of the complementary aperture field $\psi_A(x,y)$. The self-healing distance therefore is considered as a minimum distance $d$ at which the field passing through the obstruction almost fully recovers $\psi_{\B}(x,y,d)\approx \psi(x,y,d)$. From Eq.~\ref{Eq:Bessel_obstructed}, this is equivalent to $\psi_A(x,y,d)\approx0$. Therefore, our further analysis focuses on the on-axis diffraction of the complementary-aperture field $\psi_A(x,y,z)$. 

All plane-wave components on a ring $k_{r}(n)$ share the same propagation angle $\tan\theta_n=\frac{k_{r}(n)}{k_{z}(n)}$. Due to cylindrical symmetry, the self-healing distance of all plane wave components with a single $k_{r}(n)$ will be the same. Therefore, we can simply consider a single plane-wave component on the $n^{\mathrm{th}}$ ring passing through the complementary aperture, with transverse wave-vector components
\begin{equation}\label{Eq:kx_ky_n}
k_{x,n}=\kr(n)\cos\phi,
\qquad
k_{y,n}=\kr(n)\sin\phi,
\end{equation}
so that
\begin{equation}
\psi^{\mathrm{pw}}_n(x,y,z)=\exp\!\bigl[i\bigl(k_{x,n}x+k_{y,n}y+\kz(n)z\bigr)\bigr].
\end{equation}
Here $\phi$ is the propagation direction of the plane wave in the $(x,y)$ plane. 
At $z=0$ the normalized complementary-aperture field is
\begin{equation}
\psi_{A,n}(x,y,0)\propto\,\e^{i(k_{0x,n}x+k_{0y,n}y)}\tau_A(x,y).
\end{equation}
The corresponding spatial spectrum is
\begin{equation}
\widetilde{\psi}_{A,n}(k_x,k_y)=\frac{1}{2\pi}\iint \psi_{A,n}(x,y,0)\e^{-i(k_x x+k_y y)}\,\dd x\,\dd y.
\end{equation}

Indeed, there are various ways of defining $d_{n}$ depending on how close $\psi_A(r,d_{n})$ is to zero. We adopt the definition of ref.\cite{Aiello14OL} and define $z_{\mathrm{sh}}$ as the distance at which the displacement of the centroid of the beam $|\langle \textbf{r} \rangle|$ from the $z$-axis equals the half-width of the intensity distribution:
\begin{equation}\label{Eq:expectation_value}
    \rho_n(d_{n})=\langle \textbf{r}\cdot\textbf{r}\rangle_{f_n}/|\langle\textbf{r}\rangle_{f_n}|^2=2,
\end{equation}
where $\textbf{r}=\hat{x}x+\hat{y}y$, $\langle\cdot\rangle$ is the expectation value with respect to the intensity distribution $f_n(x,y,z)$ that is defined as 
\begin{equation}
    f_n(x,y,z)=\frac{|\psi_{A,n}(x,y,z)|^2}{\iint|\psi_{A,n}(x,y,z)|^2 \dd x \dd y}.
\end{equation}
To find the $\rho(z)$ it is possible to write
\begin{equation}
\expval{\zeta}_{f_n}=z\,\mu_{\zeta,n},
\qquad
\expval{\zeta^2}_{f_n}=\sigma_{\zeta,n}^2+z^2 v_{\zeta,n}^2,
\qquad \zeta\in\{x,y\},
\end{equation}
with
\begin{align}\label{Eq:mu_Bessel}
\mu_{\zeta,n}&=\iint \frac{k_{\zeta}}{\kz}\,\abs{\pdv{\widetilde{\psi}_{A,n}}{k_{\zeta}}}^2\,\dd k_x\,\dd k_y,
\\
\sigma_{\zeta,n}^2&=\iint \abs{\pdv{\widetilde{\psi}_{A,n}}{k_{\zeta}}}^2\,\dd k_x\,\dd k_y,
\\
v_{\zeta,n}^2&=\iint \frac{k_{\zeta}^2}{\kz^2}\,\abs{\widetilde{\psi}_{A,n}}^2\,\dd k_x\,\dd k_y. \label{Eq:v_Bessel}
\end{align}
Using the paraxial approximation of $k_z$ and the evenness of the shifted aperture spectrum, the odd terms vanish, and one obtains
\begin{align}
\mu_{x,n}=&v_{x,n}=\tan\theta_n\cos\phi,\\
\mu_{y,n}=&v_{y,n}=\tan\theta_n\sin\phi,
\end{align}
and
\begin{align}
v_{x,n}^2+v_{y,n}^2=\tan^2\theta_n&\equiv \xi_n^2,  \\
\sigma_{x,n}^2+\sigma_{y,n}^2 &\equiv \gamma^2.
\end{align}

Therefore, the ratio is
\begin{equation}\label{Eq:rho_Bessel}
\rho_n(z)=\frac{\sigma_{x,n}^2+\sigma_{y,n}^2+z^2 \xi_n^2}{z^2 \xi_n^2}
=1+\frac{\gamma^2}{z^2 \xi_n^2},
\end{equation}
where the geometric parameter $\gamma$ is determined by the shape of the aperture. By definition, $\rho_n(d_{n})=2$, so the self-healing distance of the Bessel beam on the $n^{\mathrm{th}}$ ring can be described by
\begin{equation}
d_{n}=\frac{\gamma}{\xi_n}.
\end{equation}

\subsection{Square and soft-Gaussian obstructions}
The geometric parameters $\gamma$ for square and soft-Gaussian obstructions have been derived in ref.\cite{Aiello14OL}, which we adopt here. For the square aperture of width $2a$, the aperture profile is described by 
\begin{equation}
\tau_{\square}(x,y)=\Thetab(a-\abs{x})\Thetab(a-\abs{y}),
\end{equation}
where $\Theta(x)$ is the Heaviside step-function. The geometric parameter $\gamma$ can be shown to take the form 
$\gamma^2_{\square}=\frac{4a^2}{\pi}$ ~\cite{Aiello14OL},
so that
\begin{equation}
d_{n}^{(\square)}=\frac{2a}{\sqrt{\pi}\,\xi_n}.
\end{equation}

For the soft-Gaussian aperture with a transmission function
\begin{equation}
\tau_G(r)=\exp\!\qty(-\frac{r^2}{2a^2}),
\end{equation}
the geometric parameter $\gamma$ can be shown to take the form $\gamma^2_G=2a^2$ \cite{Aiello14OL},
which yields
\begin{equation}
d_{n}^{(G)}=\frac{\sqrt{2}\,a}{\xi_n}.
\end{equation}

\subsection{Circular obstruction}

For a circular obstruction of radius $a$, the complementary aperture is the disk
\begin{equation}
\tau_{\circ}(r)=\Thetab(a-r).
\end{equation}
The normalized representative-wave field is then
\begin{equation}
\psi_{\circ,n}(x,y,0)=\frac{1}{\sqrt{\pi a^2}}\,\e^{i(k_{0x,n}x+k_{0y,n}y)}\Thetab(a-r),
\end{equation}
whose angular spectrum is the shifted Airy amplitude
\begin{equation}
\widetilde{\psi}_{\circ,n}(k_x,k_y)=\frac{1}{2\pi\sqrt{\pi a^2}}\iint_{r\le a}
\e^{-i[(k_x-k_{0x,n})x+(k_y-k_{0y,n})y]}\,\dd x\,\dd y
=\frac{\Jone(aQ_n)}{\sqrt{\pi}\,Q_n},
\end{equation}
with
\begin{equation}
Q_n=\sqrt{\qty(k_x-k_{0x,n})^2+\qty(k_y-k_{0y,n})^2}.
\end{equation}
Unlike the square case, no additional Gaussian surrogate is required: the second-moment coefficient can be obtained exactly from Parseval's theorem,
\begin{align}
\gamma^2_{\circ}
&=\iint \qty(\abs{\pdv{\widetilde{\psi}_{\circ,n}}{k_x}}^2+\abs{\pdv{\widetilde{\psi}_{\circ,n}}{k_y}}^2)\,\dd k_x\,\dd k_y
=\iint (x^2+y^2)\abs{\psi_{\circ,n}(x,y,0)}^2\,\dd x\,\dd y
\\
&=\frac{1}{\pi a^2}\int_0^{2\pi}\dd\varphi\int_0^a r^3\,\dd r
=\frac{a^2}{2}.
\end{align}
Therefore
\begin{equation}
\rho_n^{(\circ)}(z)=1+\frac{a^2}{2z^2 \xi_n^2},
\qquad
d_{n}^{(\circ)}=\frac{a}{\sqrt{2}\,\xi_n}.
\end{equation}

\section{Self-healing of the Montgomery pattern}

We extend the analysis to the Montgomery fields, which are a coherent superposition of Bessel beams with a \textit{non-overlapping} spatial spectrum $k_r(n)$. Assume that all representative plane waves in each Montgomery ring are aligned along the same transverse unit vector $\hat{\mathbf{u}}=(\cos\phi,\sin\phi)$, so the corresponding transverse wave-vector components can be represented by Eq. ~\ref{Eq:kx_ky_n}. 
Let $\eta_n\ge 0$ be the intensity coefficient of each ring,
\begin{equation}\label{Eq:Power_coeff}
\sum_{n=1}^{N}\eta_n=1.
\end{equation}
We can also relate this coefficient to the complex weight coefficient introduced in Eq.~\ref{Eq:Montg_spectrum} via $\eta_n\propto\frac{|\widetilde{c}_n|^2}{\sum_{1}^{N}|\widetilde{c}_n|^2}$
For thin, non-overlapping rings, cross terms in the quadratic moments are negligible. The moments of the Montgomery field are therefore the weighted sums of the moments of each ring given in Eq.~\ref{Eq:mu_Bessel}-\ref{Eq:v_Bessel},
\begin{equation}
\mu_x^{(M)}=\qty(\sum_{n=1}^{N}\eta_n \xi_n)\cos\phi,
\qquad
\mu_y^{(M)}=\qty(\sum_{n=1}^{N}\eta_n \xi_n)\sin\phi,
\end{equation}
\begin{equation}
v_x^{2(M)}+v_y^{2(M)}=\sum_{n=1}^{N}\eta_n \xi_n^2,
\qquad
\sigma_x^{2(M)}+\sigma_y^{2(M)}=\sum_{n=1}^{N}\eta_n\gamma^2.
\end{equation}
Substituting these expressions into Eq.~\ref{Eq:rho_Bessel} gives the ratio
\begin{equation}
\rho_{\M}(z)=\frac{\sum_{n=1}^{N}\eta_n\gamma^2+z^2\sum_{n=1}^{N}\eta_n \xi_n^2}{z^2\qty[(\mu_x^{(M)})^2+(\mu_y^{(M)})^2]}
=\frac{\sum_{n=1}^{N}\eta_n\gamma^2+z^2\sum_{n=1}^{N}\eta_n \xi_n^2}{z^2\qty(\sum_{n=1}^{N}\eta_n \xi_n)^2}. 
\end{equation}
Since the geometric parameter $\gamma$ is independent of $n$, and taking into account Eq.~\ref{Eq:Power_coeff}, the corresponding self-healing distance is

\begin{equation}\label{Eq:zmin_Montgomery}
    d_{\M}=hz_s, \qquad h=\left\lceil\frac{\gamma}{\vartheta\zs} \right\rceil
\end{equation}

The geometric parameter $\gamma$ in the numerator of Eq.~\ref{Eq:zmin_Montgomery} is determined by the shape of the aperture, threshold value for self-healing, and is the same as for the Bessel beam. Meanwhile, the denominator $\vartheta=\sqrt{2\qty(\sum_{n=1}^{N}\eta_n \xi_n)^2-\sum_{n=1}^{N}\eta_n \xi_n^2}$ is dictated by the spatial distribution $k_r(n)$ in Eq.~\ref{Eq:MontCondition} and the complex amplitude distribution $\widetilde{c}_n$. 
\bibliography{Montgomery}